\documentclass[
    ,final            
  ]
  {aipproc}

\layoutstyle{6x9}

\newcommand{\Dslash}{\rlap{/}\kern-2.0pt D}

\def\mres{m_{\rm res}}
\def\GeV{\rm GeV}
\def\MeV{\rm MeV}
\def\LorInd{{\mu_1\mu_2\cdot\cdot\cdot\mu_n}}
\def\LorIndtwo{{\mu_2\cdot\cdot\cdot\mu_n}}
\def\Dcc{\stackrel{\,\leftrightarrow}{D}}

\begin{document}

\title{Spin on the lattice}

\author{Konstantinos Orginos\footnote{
For the RBC collaboration. The current members of the RBC collaboration are: Y.~Aoki, T.~Blum, N.~Christ, M.~Creutz, C.~Dawson, T.~Izubuchi, L.~Levkova, X.~Liao, G.~Liu, R.~Mawhinney, Y.~Nemoto, J.~Noaki, S.~Ohta, K.~Orginos, S. Prelovsek, S.~Sasaki and A.~Soni. Plenary talk presented at SPIN2002}}{
  address={RIKEN-BNL Research Center, Brookhaven National Laboratory, Upton, NY 11973, USA}
}



\begin{abstract}
 I review the current status of hadronic structure computations on the
lattice.  I describe the basic lattice techniques and difficulties 
and present some of the latest lattice results; in particular recent results
of the RBC group using domain wall fermions are also discussed.
\end{abstract}

\maketitle
\vspace{-.5cm}


 Understanding the basic properties of matter requires the
understanding of the nucleon structure. Quantum Chromodynamics (QCD)
is the theory describing strong interactions and hence is responsible
for the properties of the nuclear matter. Although QCD has been
around for more than twenty years, its non-perturbative nature
is an obstacle to the direct connection of low energy physics to
quarks and gluons, the fundamental degrees of freedom of the
theory. Unlike QED, non-perturbative techniques had to be developed in
order to understand the QCD predictions at low energies. The lattice
formulation of QCD is both a non-perturbative way to define the theory
and a very powerful tool in understanding its properties.

Deep inelastic scattering of leptons on nucleons has been an important
tool in understanding the structure of hadrons. Over the last few
decades experiments at SLAC, Fermilab, CERN, DESY, and more recently
at RHIC and JLAB, have measured the quark and gluon light cone
distribution functions of the nucleon.
These experiments have substantially advanced our knowledge of
the properties of hadrons. However, we would also like to study
how this observed rich phenomenology arises form first principles,
i.e. QCD. With todays advances in computer technology, algorithms,
and recent developments in lattice regularization of
fermions, lattice calculations can complement the experimental effort
and promote our understanding of the non-perturbative nature of QCD.
\vspace{-1cm}

\subsection{The Lattice Formulation}
The continuum Euclidean path integral can be defined using the lattice
regulator~\cite{Wilson:1974sk}. In order to preserve gauge invariance
the lattice gauge fields are link variables 
\begin{equation}
U_\mu(x)=e^{i\int_x^{x+\hat\mu} d\tau A_\mu(\tau)},  
\end{equation}
where $A_\mu$ are the continuum gauge fields. The fermion fields
live on the sites of the lattice. Naive discretization of the
continuum fermionic action leads to the so-called fermion doubling
problem. This problem can be avoided by either reinterpreting the
additional light fermions as extra flavors (the Kogut-Susskind approach)
or by introducing an irrelevant dimension 5 operator that breaks
chiral symmetry on the lattice and gives mass proportional to the
inverse lattice cutoff to the fermion doublers (the Wilson approach). Recently,
new lattice fermionic actions that both preserve chiral symmetry on the lattice
and do not suffer from the fermion doubling problem have been introduced.
Such fermionic actions are the domain wall
fermions~\cite{Kaplan:1992bt,Kaplan:1993sg,Shamir:1993zy,Furman:1995ky},
the overlap fermions~\cite{Narayanan:1994sk}, and the fixed point
fermions~\cite{DeGrand:1995ji,Bietenholz:1996cy,Hasenfratz:2000xz}.
Having defined the lattice theory, correlation functions
can be evaluated using  Monte-Carlo integration in Euclidean space.

However, parton distribution functions are defined in the Minkowski
space, and hence cannot be directly computed in lattice QCD. Using
the operator product expansion we can relate moments of the structure
functions to forward matrix elements of gauge invariant local operators
(for a pedagogical review see~\cite{Manohar:1992tz}).
These matrix elements can then be computed using lattice QCD.

\begin{figure}[t]
\includegraphics[width=5cm]{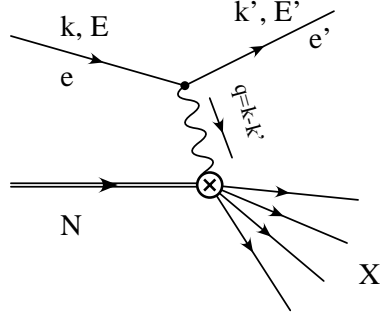}
\vspace{-1cm}
\caption{Deep inelastic scattering.}
\label{fig:DIS}
\vspace{-4cm}
\end{figure}

In a deep inelastic process (see Fig.~\ref{fig:DIS}) the cross section is
given by 
\begin{equation}
\frac{d^2\sigma}{d\Omega dE'} = \frac{1}{2 m_N} \frac{\alpha^2}{q^4} 
\frac{E'}{E} l^{\mu\nu} W_{\mu\nu}
\end{equation}
where $ l^{\mu\nu}$ is the lepton tensor, $W_{\mu\nu}$ is the
hadronic tensor, $q$ is the momentum transfer, $m_N$ is the nucleon
mass. The initial and final energy and momentum of the lepton are
  ($E,k$) and ($E',k'$) respectively.

The hadronic tensor can be decomposed in the symmetric $W^{\{\mu\nu\}}$
and anti-symmetric  $W^{[\mu\nu]}$ pieces:

\begin{equation}
W^{\mu\nu} =    W^{[\mu\nu]} +  W^{\{\mu\nu\}} 
\end{equation}

The symmetric piece defines the unpolarized structure functions $F_1$
and $F_2$ ($F_3$ also for neutrino scattering).
\begin{equation}
W^{\{\mu\nu\}}(x,Q^2) =  \left(-g^{\mu\nu} + \frac{q^\mu q^\nu}{q^2}\right)
 F_1(x,Q^2) \nonumber + \!
\left(p^\mu-\frac{\nu}{q^2}q^\mu\right)\!\!\!
\left(p^\nu-\frac{\nu}{q^2}q^\nu\right)\!
\frac{F_2(x,Q^2)}{\nu},
\end{equation}
while the anti-symmetric defines the polarized structure functions $g_1$
and $g_2$
\begin{equation}
W^{[\mu\nu]}(x,Q^2) = i\epsilon^{\mu\nu\rho\sigma} q_\rho
\left(\frac{s_\sigma}{\nu}(g_1(x,Q^2)+g_2(x,Q^2)) - 
\frac{q\cdot s p_\sigma}{\nu^2}{g_2(x,Q^2)} \right).
\end{equation}

where $p_\mu$ and $s_\mu$ are the nucleon momentum and spin vectors,
$\nu = q\cdot p$, $s^2 = -m_N^2$, $x=Q^2/2\nu$  and $Q^2=-q^2$.

At the leading twist in the operator product expansion the moments
of the structure functions can be factorized at a scale $\mu$ in
hard perturbative contributions (the Wilson coefficients)
and low energy matrix elements of local gauge invariant operators:
\begin{eqnarray}
2 \int_0^1 dx x^{n-1} {F_1(x,Q^2)} 
&=& \sum_{q=u,d} c^{(q)}_{1,n}(\mu^2/Q^2,g(\mu))\: v_n^{(q)}(\mu),
\nonumber \\ 
\int_0^1 dx x^{n-2} {F_2(x,Q^2)} 
&=& \sum_{q=u,d} c^{(q)}_{2,n}(\mu^2/Q^2,g(\mu))\: v_n^{(q)}(\mu), 
\nonumber \\
2\int_0^1 dx x^n {g_1(x,Q^2)} 
  &=& \frac{1}{2}
\sum_{q=u,d} e^{(q)}_{1,n}(\mu^2/Q^2,g(\mu))\: a_n^{(q)}(\mu),\nonumber
 \\
2\int_0^1 dx x^n {g_2(x,Q^2)}
  &=& \frac{1}{2}\frac{n}{n+1} \sum_{q=u,d} [e^{(q)}_{2,n}(\mu^2/Q^2,g(\mu))
\: {d_n^{(q)}(\mu)} - \nonumber \\
  &-&  e^{(q)}_{1,n}(\mu^2/Q^2,g(\mu))\: a_n^{(q)}(\mu)]
\label{eq:Moments}
\end{eqnarray}
where $c^{(q)}_{i,n},e^{(q)}_{i,n}$ are the Wilson coefficients and
$ v_n^{(q)}(\mu), a_n^{(q)}, {d_n^{(q)}(\mu)}$ are the non-perturbative
matrix elements. At the leading twist $ v_n^{(q)}(\mu)$ and $a_n^{(q)}$
are related to the parton model distribution functions 
$\langle x^n\rangle_q$ and $\langle x^n\rangle_{\Delta q}$:
\begin{eqnarray}  
  \langle x^{n-1}\rangle_q = v_n^{(q)} \;\;\;\;\;\;\;\;\;\;\;\;\;\;\;
  \langle x^{n}\rangle_{\Delta q} = \frac{1}{2} a_n^{(q)}
\end{eqnarray}

In order to extract $ v_n^{(q)}(\mu), a_n^{(q)}$, and ${d_n^{(q)}(\mu)}$
we need to compute non-perturbatively  the following matrix elements: 
\begin{eqnarray}
 \frac{1}{2} \sum_s \langle p,s|{{\cal O}^{q}_{\{\LorInd\}}}
 |p,s\rangle &=&
 2 v_n^{(q)}(\mu)\times
 [ p_{\mu_1}p_{\mu_2}
\cdot\cdot\cdot p_{\mu_n}+
\cdot\cdot\cdot -tr]\nonumber\\
 -\langle p,s|{{\cal O}^{5q}_{\{\sigma\LorInd\}}} |p,s\rangle &=&
 \frac{1}{n+1}a_n^{(q)}(\mu)\times
 [ s_\sigma p_{\mu_1}p_{\mu_2}\cdot\cdot\cdot p_{\mu_n}+\cdot\cdot\cdot
 -tr]\nonumber\\
 \langle p,s|{\cal O}^{[5]q}_{[\sigma\{\mu_1]\LorIndtwo\}}
 |p,s\rangle &=&
 \frac{1}{n+1}{d_n^{ q}}(\mu)\times
 [ (s_\sigma p_{\mu_1} - s_{\mu_1} p_{\sigma})p_{\mu_2}
\cdot\cdot\cdot p_{\mu_n}+\cdot\cdot\cdot -tr]\nonumber\\
\label{eq:matel}
\end{eqnarray}
$\{\}$ implies symmetrization and $[]$ anti-symmetrization of indices.
The nucleon states $|p,s\rangle$ are normalized so that
$\langle p,s|p',s'\rangle = (2\pi)^3 2E(p)\delta(p-p')\delta_{s,s'}$ and
$s^2=-m^2_N$. The
operators $\cal O$ are
\begin{eqnarray}
  {\cal O}^q_\LorInd  &=&
 \left(\frac{i}{2}\right)^{n-1}\bar{q}\gamma_{\mu_1} 
\Dcc_{\mu_2}\cdot\cdot\cdot \Dcc_{\mu_n}q -trace\nonumber\\
{\cal O}^{5q}_{\sigma\LorInd} &=&
 \left(\frac{i}{2}\right)^{n}\bar{q}\gamma_{\sigma}\gamma_5 
 \Dcc_{\mu_2}\cdot\cdot\cdot \Dcc_{\mu_n}q -trace
\label{eq:upol_pol_ops}
\end{eqnarray}
where $\Dcc = \overrightarrow{D} - \overleftarrow{D}$ and $\overrightarrow{D}$,
$\overleftarrow{D}$ are covariant derivatives acting on the right and the
left respectively.

In Drell-Yan processes the transversity distribution
$\langle x \rangle_{\delta q}$ can be measured 
(for details see~\cite{Jaffe:1991kp,Jaffe:1992ra,Barone:2001sp}).
The relevant matrix element is 
\begin{equation}
  \langle p,s|{{\cal O}^{\sigma q}_{\rho\nu\{\LorInd\}}}
 |p,s\rangle =
 \frac{2}{m_N}{\langle x^{n}\rangle_{\delta q}}(\mu)\times
 [(s_\rho p_{\nu} - s_{\nu} p_\rho)p_{\mu_1}p_{\mu_2}\cdot\cdot\cdot p_{\mu_n}
+\cdot\cdot\cdot -tr]\nonumber
\label{eq:matel_trans}
\end{equation}
and the operators ${\cal O}^{\sigma q}$ are
\begin{equation}
{{\cal O}^{\sigma q}_{\rho\nu\LorInd}} =
\left(\frac{i}{2}\right)^{n}\bar{q}\gamma_5 \sigma_{\rho\nu} \Dcc_{\mu_1}\cdot\cdot\cdot \Dcc_{\mu_n}q - trace.
\label{eq:trans_ops}
\end{equation}

\subsection{Lattice matrix elements}
In order to calculate on the lattice the needed matrix elements
we have   to compute nucleon three point functions
\begin{equation}
  C^\Gamma_{3pt}(\vec{p},t,\tau) = \sum_{\alpha,\beta}\Gamma^{\alpha,\beta}\langle J_{\beta}(\vec{p},t) {\cal  O}(\tau)\bar{J}_\alpha(\vec{p},0) \rangle
\label{eq:3pt}
\end{equation}
and nucleon two point functions
\begin{equation}
  C_{2pt}(\vec{p},t) = \sum_{\alpha,\beta}
\left.\frac{1+\gamma_4}{2}\right|_{\alpha,\beta}
\langle J_{\beta}(\vec{p},t) \bar{J}_\alpha(\vec{p},0) \rangle
\end{equation}
where $\bar{J}(\vec{p},0)$ and $J(\vec{p},t)$
 are creation and annihilation operators
of states with the quantum numbers of the nucleon. For unpolarized
matrix elements $\Gamma=\frac{1+\gamma_4}{2}$ while for the polarized
 $\Gamma=\frac{1+\gamma_4}{2}i\gamma_5\gamma_k$ ($k\ne 4$). The 
$\frac{1+\gamma_4}{2}$ factor is  for projecting out the positive parity
part of the baryon propagator i.e. the nucleon.
For the proton a typical choice is
\begin{equation}
  J_\alpha(\vec{p},t) = \sum_{\vec{x},a,b,c} e^{-i\vec{p}\cdot\vec{x}}\epsilon^{abc} \left[u^a(x,t) C \gamma_5 d^b(x,t)\right] u^c_\alpha(x,t)
\end{equation}
where $C=\gamma_4\gamma_2$ the charge conjugation matrix, $\alpha$ is a spinor
index and $a,b,c$ are color indices.
When $t\gg \tau \gg 0$ 
\begin{eqnarray}
 C_{2pt}(\vec{p},t) &=& Z_N \frac{E_N(\vec{p})+m_N}{E(\vec{p})}
 e^{-E_N(\vec{p})t} + \cdots \nonumber\\
  C^\Gamma_{3pt}(\vec{p},t,\tau) &=& Z_N \sum_{\alpha,\beta,s}
\Gamma_{\alpha\beta}U^\alpha_N(p,s) 
              \langle p,s | {\cal  O} |p,s \rangle \bar{U}^\beta_N(p,s) 
	         e^{- E_N(\vec{p})t} + \cdots
\label{eq:asympt_3pt_2pt}
\end{eqnarray}
where $U(p,s)$ is the nucleon spinor which satisfies the Dirac equation and 
$\langle 0| J_\alpha(\vec{p},t)|p,s\rangle = \sqrt{Z_N} U^\alpha(p,s)$.
From Eq.~\ref{eq:asympt_3pt_2pt} and  Eq.~\ref{eq:matel} (or Eq.~\ref{eq:matel_trans}) the required matrix elements can be extracted from the ratio
of three point functions over two point functions. In practice we would like
to achieve the asymptotic behavior of Eq.~\ref{eq:asympt_3pt_2pt}
with as small  as possible $t$ and $\tau$. For that reason the
interpolating operator $J$ is tuned so that the overlap with the
exited nucleon states would be as small as possible.
 For more details on the technical aspects
of the lattice calculation the reader may refer 
to~\cite{Martinelli:1989rr,Gockeler:1996wg,Gockeler:2000ja,Dolgov:2002zm}.

In order to reduce the computational cost of calculating the above
correlation functions some times the so-called quenched approximation
is used. In this approximation the quark loop contributions to the
path integral are ignored. Quenching reduces the computational cost by
several orders of magnitude, while for certain quantities it
introduces a systematic error $\sim 10\%$.\footnote{Note that there
are quantities for which the quenched approximation introduces
uncontrollable errors.}  In addition, lattice computations are
typically performed with heavier quark masses than the physical up and
down quarks. Hence we have to perform extrapolations to the chiral
limit. If the quark masses are light enough, chiral perturbation
theory~\cite{Arndt:2001ye,Chen:2001eg,Chen:2001gr} can be used to
calculate the dependence of the matrix elements on the quark
mass.\footnote{In the case of the quenched approximation the so-called
quenched chiral perturbation theory is used.} Finally the lattice
matrix elements have to be renormalized, typically to $\overline{\rm
MS}$, and extrapolated to the continuum limit.
\vspace{-.5cm}
\subsection{Renormalization}
The renormalized operators at scale $\mu$ are obtained from the 
lattice operators calculated at lattice spacing $a$ from
\begin{equation}
  {\cal O}^{ren}(\mu) = Z(\mu;a) {\cal O}^{lat}(a)
\label{eq:renorm_mult}
\end{equation}
in the case of multiplicatively renormalized operators. In general,
there is operator mixing and as a result  the above relation becomes
\begin{equation}
  {\cal O}_i^{ren}(\mu) =  Z(\mu;a)\left[{\cal O}^{lat}_i(a) +
\sum_{j\ne i} a^{d_j - d_i}Z_{ij}(\mu;a) {\cal O}^{lat}_j(a)\right],
\label{eq:renorm_gen}
\end{equation}
where ${\cal O}_j$ are a set of operators allowed by symmetries to mix,
and $d_j$ is the dimension of each operator. It is evident that if
mixing with lower dimensional operators occur, the mixing coefficients are
power divergent as we approach the continuum limit.
 Hence we have to compute these terms non-perturbatively in order 
to accurately renormalize the operators. Higher dimensional
operators are typically ignored since their effects vanish in
the continuum limit. In certain cases we may want to compute
these coefficients in order to remove part of the systematic
error introduced by the finite cutoff.

The mixing of lattice operators is more complicated than that
of the continuum operators, since on the lattice we do not have all
the continuum symmetries. In particular, $O(4)$ rotational symmetry  
in Euclidean space  is broken down to the hypercubic group $H(4)$.
As a result, an irreducible representation of $O(4)$ is reducible under
$H(4)$ and hence mixing of operators that would not occur in the continuum
can occur on the lattice. For a detailed analysis
of the $H(4)$ group representations see~\cite{Mandula:1983us,Gockeler:1996mu}
 and references therein.
In lattice calculations we have
to select  carefully the lattice operators so that mixing with lower
dimensional operators does not occur and hence no power
divergent coefficients in Eq.~\ref{eq:renorm_gen} are encountered.
This turns out to be a significant constraint on how many moments can 
be practically computed on the lattice. 

Another symmetry that is broken on the lattice for Wilson fermions is
chiral symmetry. This  results in mixings with lower dimensional
operators for the $d_n$ matrix elements. Fortunately, in this case we
can use lattice fermions, such as domain wall or overlap and fixed point
fermions, that respect chiral symmetry on the lattice. 
For Wilson fermions,
 the renormalization of $d_2$ 
has been done non-perturbatively as described in~\cite{Gockeler:2000ja}.

The renormalization constants for all the operators 
relevant to structure function calculations have been computed perturbatively
for Wilson fermions, improved 
and unimproved~\cite{Capitani:1995qn,Beccarini:1995iv,Capitani:2000xi}.
Moreover, the RI-MOM scheme has been
used to renormalize non-perturbatively both local~\cite{Martinelli:1995ty}
 and derivative operators~\cite{Gockeler:1998ye,Capitani:2001em}.
In the Schroedinger functional scheme (developed by the ALPHA collaboration),
 all local operator
renormalizations and the renormalization of $v_2$ have 
been computed~\cite{Guagnelli:1999wp,Guagnelli:2000em}.
In addition, work is underway for computing the constants for
flavor singlet operators~\cite{Palombi:2002gw}. 
For domain wall fermions, 
all local operators have been renormalized 
non-perturbatively~\cite{Blum:2001sr} using the RI-MOM scheme,
and also perturbatively~\cite{Aoki:1999ky}.
\vspace{-.5cm}
\section{Lattice results}
 In the last several years, the lattice community (QCDSF/UKQCD and
LHP/SESAM collaborations) has made a substantial effort to compute the
first few moments of the nucleon structure functions.  Apart from the
constraints imposed by the renormalization of the operators mentioned
above, the requirement of having nucleon states
with non-zero momentum\footnote{operators with more than one derivative need non-zero momentum nucleon states 
see Eq.~\ref{eq:matel} and Eq.~\ref{eq:matel_trans}} limits the
number of moments we can compute. These are
the first three moments of the  unpolarized structure functions,
the first two moments of the polarized structure functions, and the
first two moments of the transversity. These computations have been
performed both in quenched and in full QCD with improved and unimproved
Wilson
fermions~\cite{Gockeler:1996wg,Gockeler:1997zr,Gockeler:2000ja,Dolgov:2002zm}.
\begin{figure}[t]
\includegraphics[width=.5\textwidth]{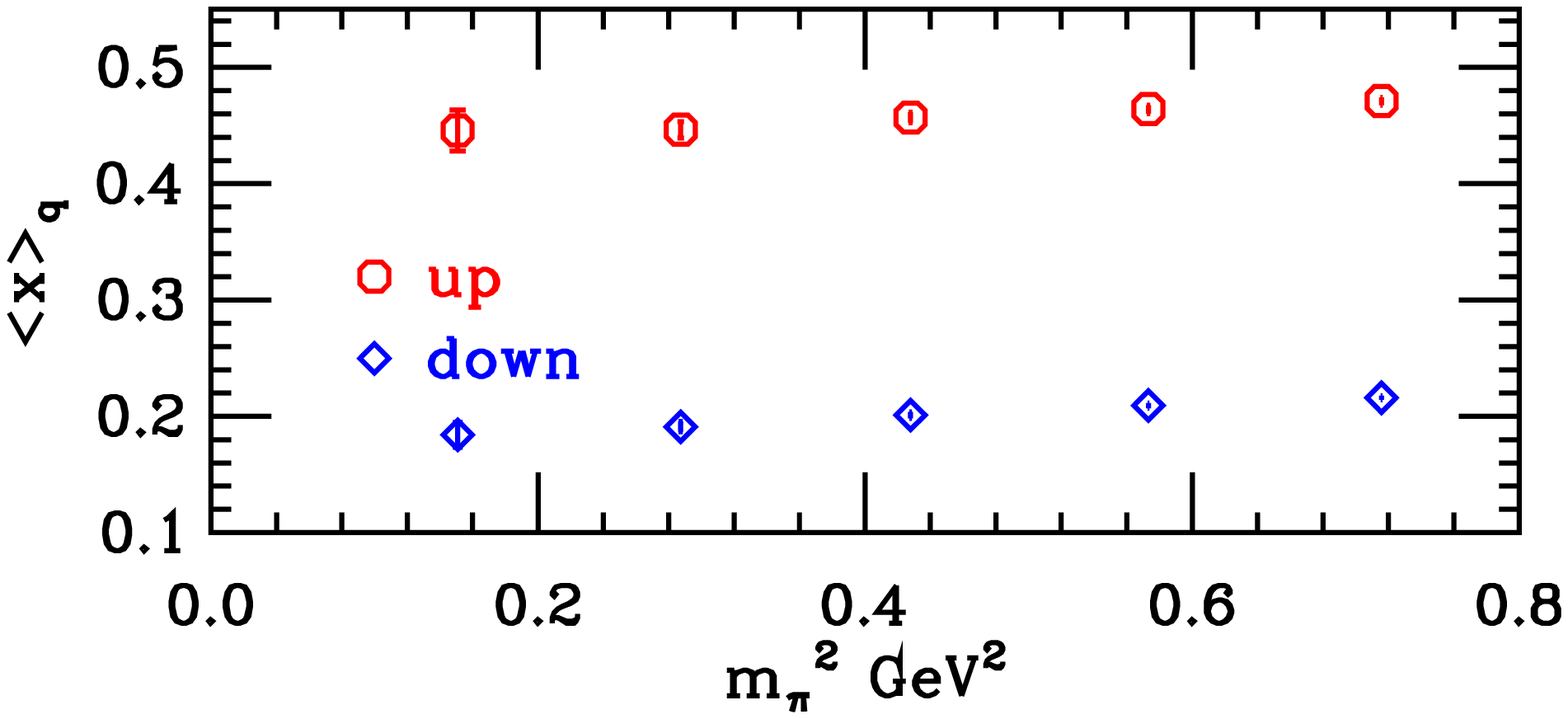}
\includegraphics[width=.5\textwidth]{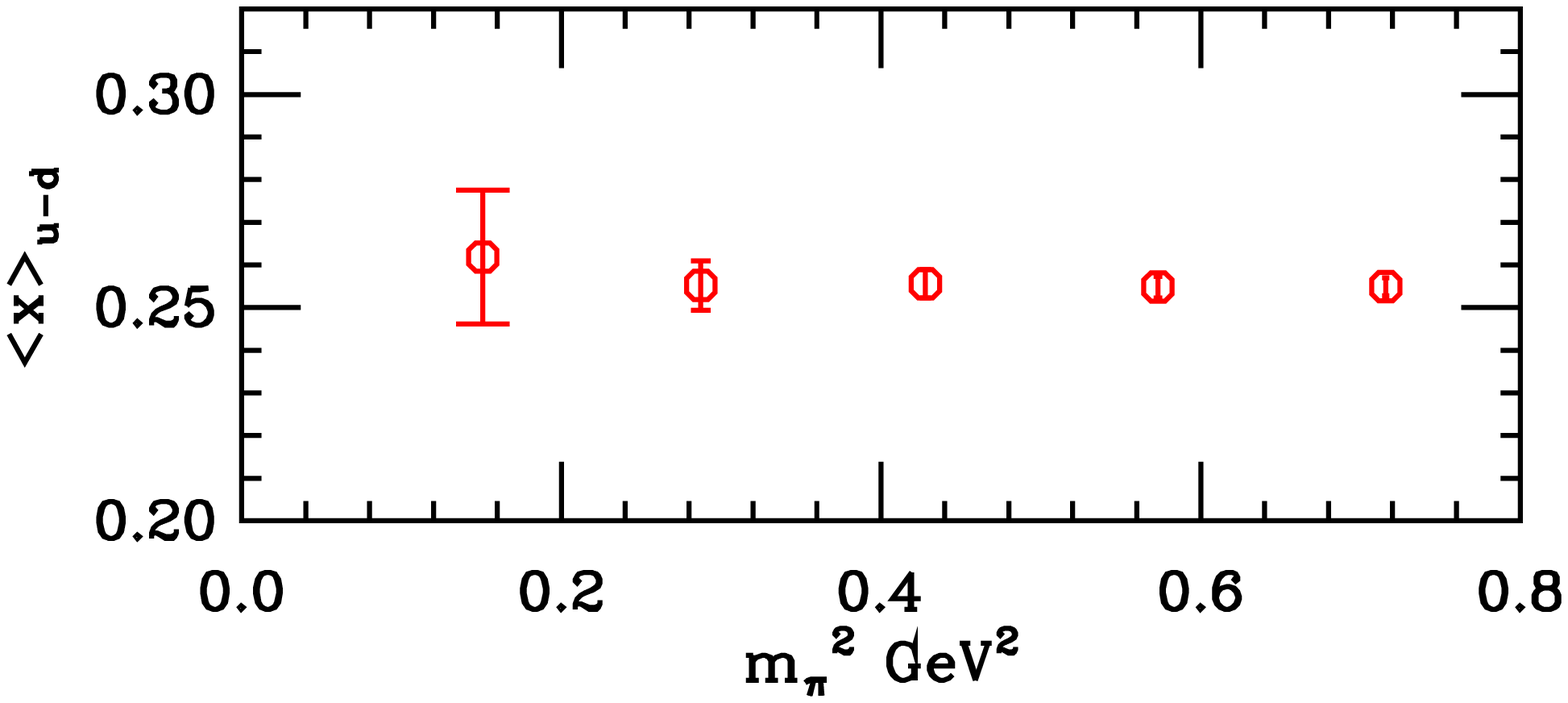}
\vspace{-.5cm}
\caption{Quark density $\langle x \rangle_q$ vs. the pion mass squared. 
{\bf[left]} The connected up (octagons) and down (diamonds)
 quark contributions. 
{\bf[right]} The flavor non-singlet $\langle x \rangle_{u-d}$.}
\label{fig:X_ns_mpi2}
\vspace{-2cm}
\end{figure}

The RBC group has recently begun quenched computations with
domain wall fermions~\cite{Orginos:2002mn}. 
Our current results are restricted only to those matrix elements that
can be computed with zero momentum nucleon states. We use the DBW2
gauge action which is known to improve the domain wall fermion chiral
properties~\cite{Orginos:2001xa,Aoki:2001dx}. We have 416 
lattices of size 
$16^3\times 32$ at $\beta=0.870$ with lattice spacing
$a^{-1}=1.3\GeV$,  providing us with a physical volume ($\sim
(2.4fm)^3$) large enough to reduce finite size effects known to affect
some nucleon matrix elements, such as
$g_A$~\cite{Sasaki:2001th,Ohta:2002ns}.  Using fifth dimension length
$L_s=16$ we achieve a residual mass $\mres \sim
0.8\MeV$~\cite{Orginos:2001xa,Aoki:2001dx}.  The input quark masses ranged
from $0.02$ to $0.10$,  providing pion masses
ranging from $390\MeV$ to $850\MeV$.  Further technical details of
our calculation can be found in~\cite{Orginos:2002mn}.
\vspace{-.5cm}

\subsection{Unpolarized Structure Functions}
The first three moments  of the unpolarized 
structure functions have been computed by QCDSF  
in the quenched approximation. The needed chiral and continuum extrapolations
have also been performed.
A summary of recent results can be found in~\cite{Gockeler:2002ek}.
In comparison with MRS phenomenological results, the lattice results
are typically higher. Also, $v_3$ is smaller than $v_4$, while 
$v_3>v_4$ is phenomenologically expected. The same computations
have been performed by LHP/SESAM in full QCD~\cite{Dolgov:2002zm}.
Their results indicate that dynamical fermions have only
a small effect on the matrix elements they studied.

It has been argued
that the main reason for such discrepancies is the fact
that lattice computations are performed at rather heavy quark
masses and then extrapolated linearly to the chiral 
limit~\cite{Detmold:2001jb,Detmold:2002nf,Thomas:2002sj}.
For that reason, we need computations at much lighter quark masses
in order to see whether there is a disagreement with  phenomenological
expectations. In quenched QCD, a study with very light quark masses
has been done~\cite{Gockeler:2002mk} indicating that the
linear behavior persists down to 300MeV pion masses.

In Fig.~\ref{fig:X_ns_mpi2} we present our results 
for the quark density distribution $\langle x \rangle_q$ ($v_2$). 
 We plot the unrenormalized result for $\langle x\rangle_u$, 
$\langle x \rangle_d$ and the flavor non-singlet $\langle
x \rangle_{u-d}$.  Down to 380MeV pion mass
no significant curvature within our statistical 
errors can be seen.\footnote{In~\cite{Orginos:2001xa} we had an indication of
some curvature but this effect went away as we doubled the statistics.}
The ratio
${\langle x \rangle_u}/{\langle x \rangle_d}$ is $2.41(4)$, linearly
 extrapolated  to the
chiral limit, is in agreement with the quenched Wilson fermion
results~\cite{Gockeler:1996wg,Dolgov:2002zm}.
\vspace{-.5cm}
\subsection{Polarized Structure Functions}
\begin{figure}[t]
\includegraphics[width=.5\textwidth]{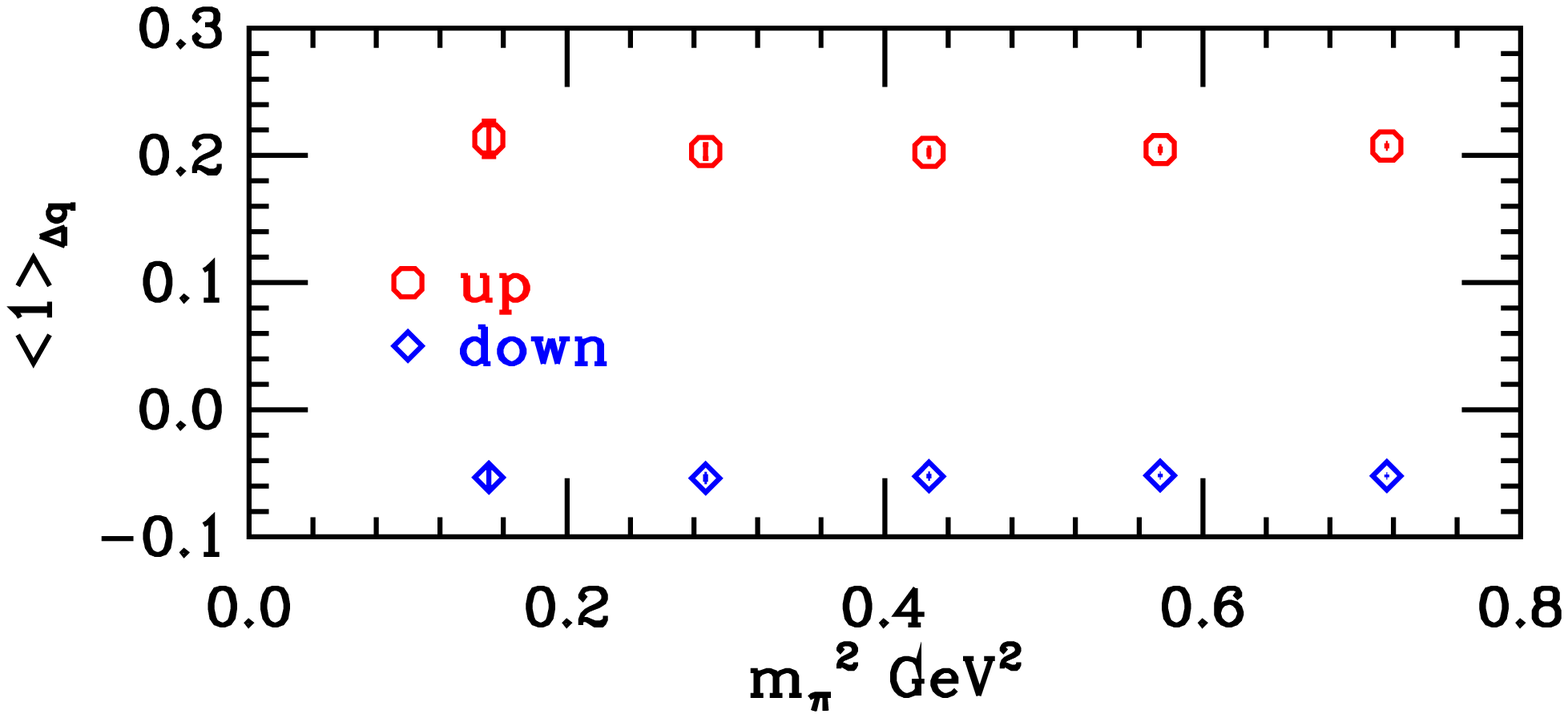}
\includegraphics[width=.5\textwidth]{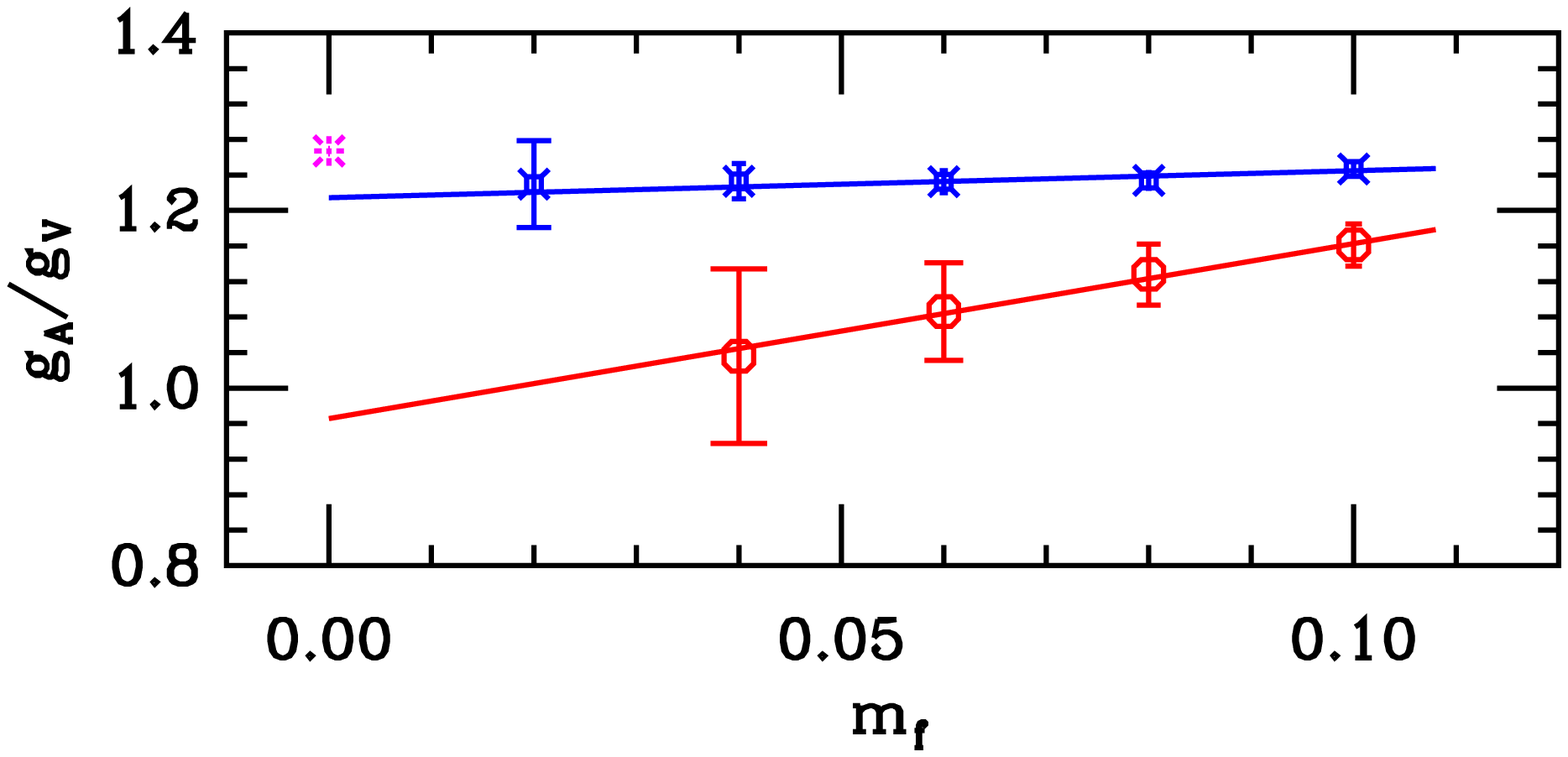}
\vspace{-.5cm}
\caption{ Helicity $\langle 1 \rangle_{\Delta q}$ vs. the pion mass squared. 
{\bf[left]}  The connected up (octagons)
 and down (diamonds) quark contributions. 
{\bf[right]} The nucleon axial charge $g_A$ i.e.
 flavor non-singlet $\langle 1 \rangle_{\Delta u- \Delta d}$.}
\label{fig:gA}
\vspace{-2cm}
\end{figure}
\begin{figure}[t]
\includegraphics[width=.5\textwidth]{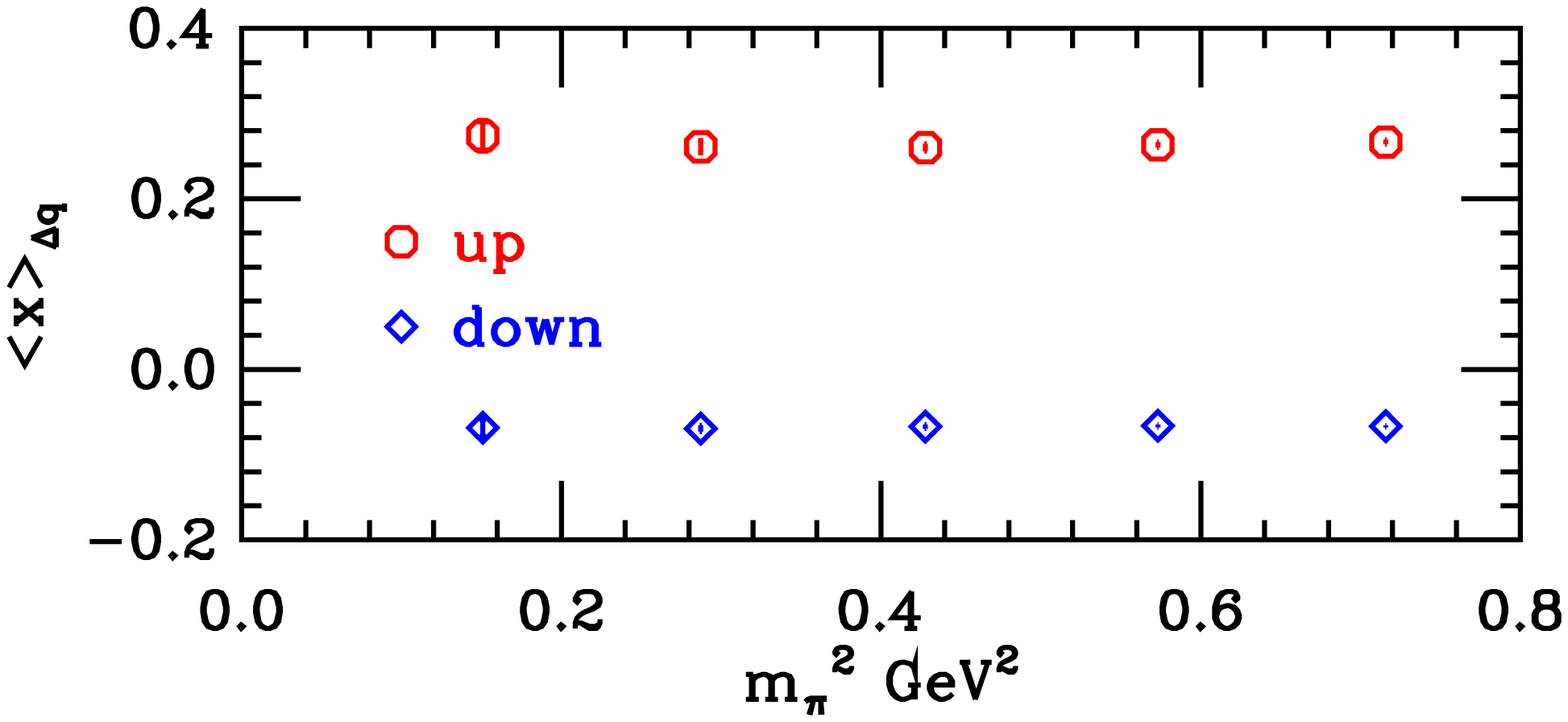}
\includegraphics[width=.5\textwidth]{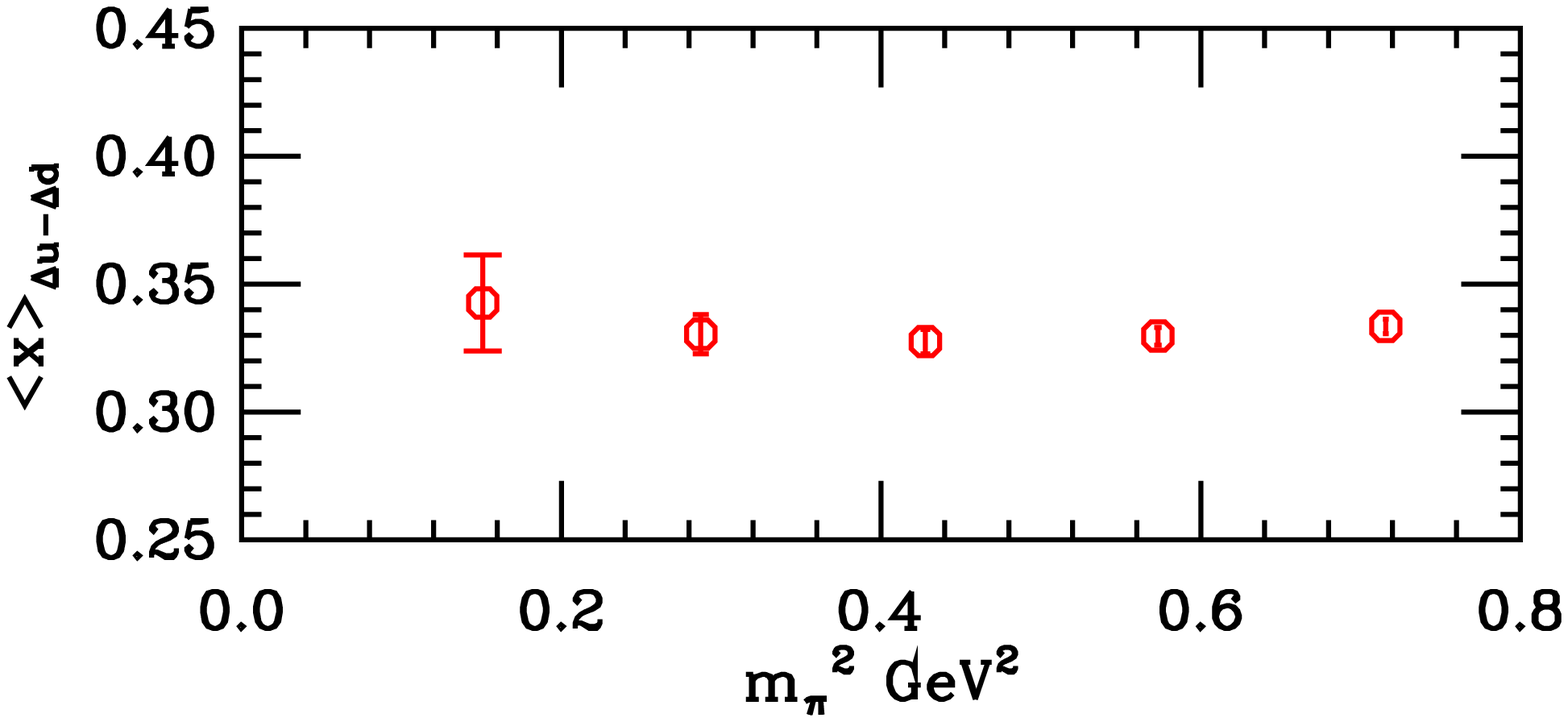}
\vspace{-.5cm}
\caption{Helicity  $\langle x \rangle_{\Delta q}$ vs. the pion mass squared. 
{\bf[left]} The connected up (octagons) and down (diamonds) quark 
contributions. {\bf[right]} The flavor non-singlet 
$\langle x \rangle_{\Delta u- \Delta d}$. }
\label{fig:XDq_vs_mpi2}
\vspace{-2cm}
\end{figure}
The nucleon axial charge $g_A$  is related 
to the first moment of the polarized structure function $g_1$.  
The current experimental  value for $g_A/g_V$ measured from neutron 
beta decays is 1.2670(30)~\cite{Hagiwara:2002pw}. 
Lattice calculations, quenched and dynamical,
 have been underestimating this quantity
typically by 10\% to 20\%~\cite{Fukugita:1995fh,Gockeler:1996wg,Gusken:1999as,Gockeler:2000ja,Dolgov:2002zm,Gockeler:2002ek}. 
For earlier calculations see also~\cite{Liu:1994ab,Dong:1995rx}.

One of the systematic errors believed to affect these calculations
is the finite volume. In order to study this effect we performed
two calculations. One with spatial volume $2.4^3fm^3$ and another
with spatial volume  $1.2^3fm^3$. Our results are shown in 
Fig.~\ref{fig:gA}[right]. Between these two volumes it is clear that
there is a finite volume effect of about 20\% at the chiral limit.
In addition, the linearly extrapolated to the chiral limit value for $g_A/g_V$ 
is 1.21(2). For a detailed analysis of this computation see~\cite{Ohta:2002ns}.
Note that for domain wall fermions $g_A/g_V$ does not
require renormalization, since the finite
renormalization constants of the axial and the vector currents $Z_A$, $Z_V$
are equal~\cite{Blum:2001sr,Dawson:2002nr}. In Fig.~\ref{fig:gA}[left]
we present the up and down quark contributions of
$\langle 1 \rangle_{\Delta q}$ for the proton renormalized using 
$Z_A = .77759(45)$~\cite{Aoki:2001dx}. In Fig.~\ref{fig:XDq_vs_mpi2} 
we present our unrenormalized data for $\langle x \rangle_{\Delta q}$. 
The ratio ${\langle x \rangle_{\Delta u}}/{\langle x \rangle_{\Delta d}}$
linearly extrapolated to the chiral limit   is roughly $-4$,
 consistent with other lattice  results~\cite{Gockeler:2000ja,Dolgov:2002zm}.
\begin{figure}[t]
\includegraphics[width=.5\textwidth]{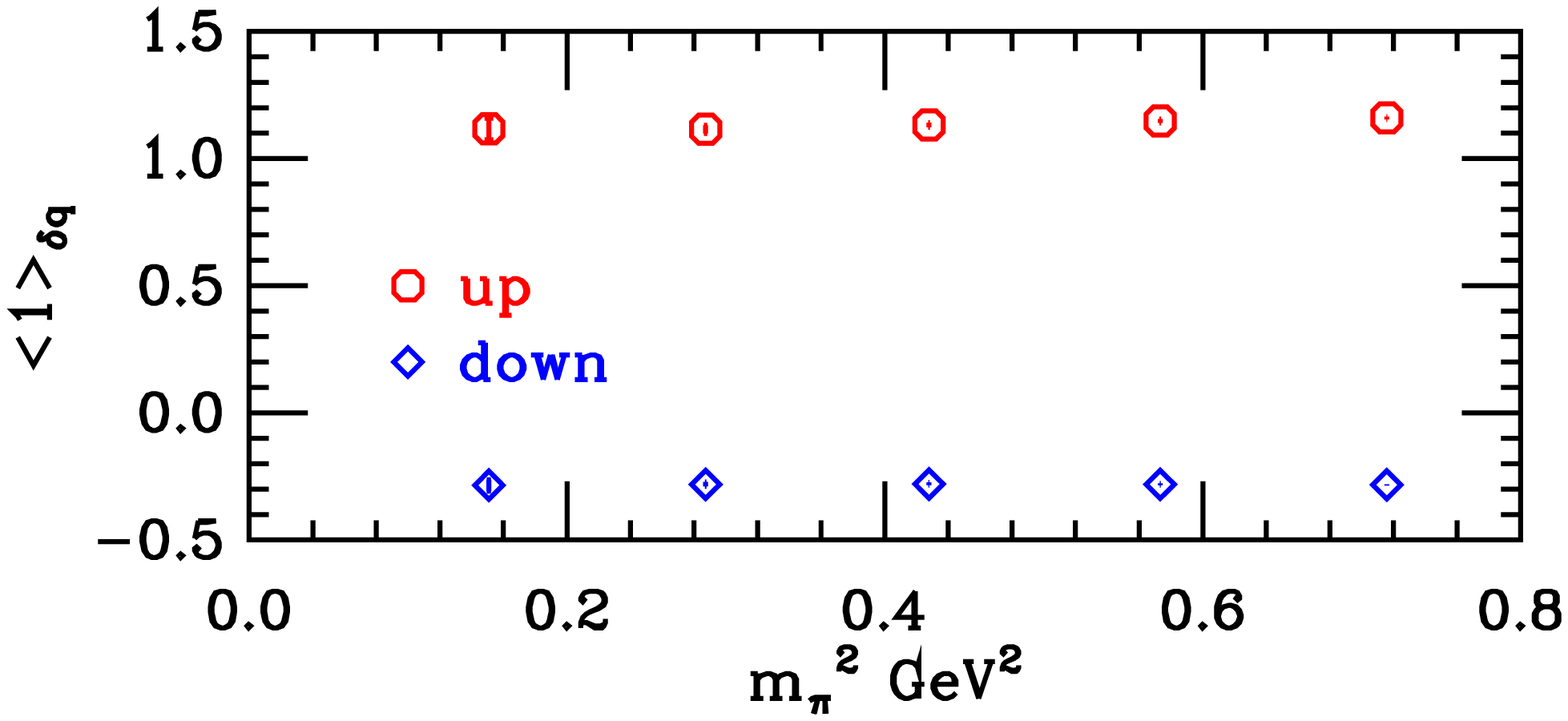}
\includegraphics[width=.5\textwidth]{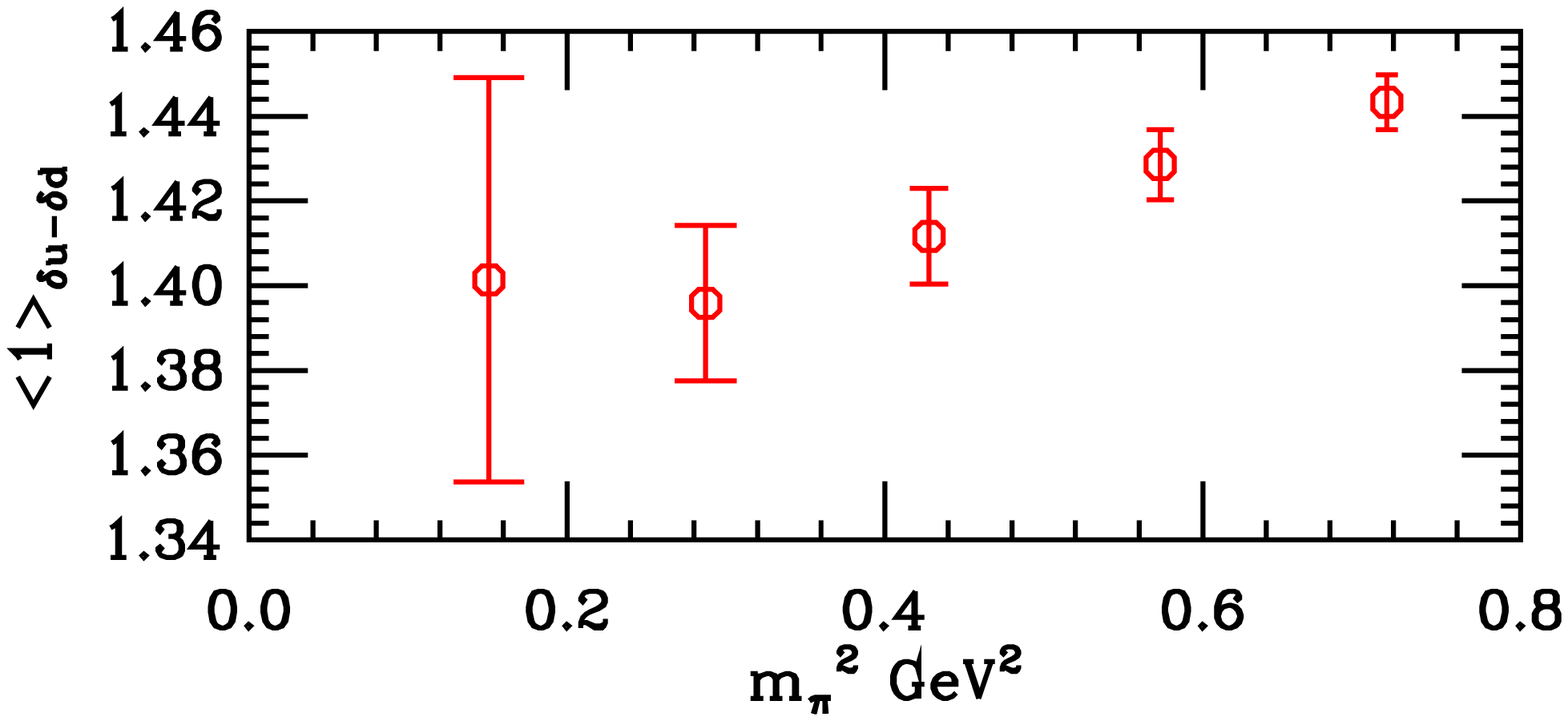}
\vspace{-.5cm}
\caption{Transversity  $\langle 1 \rangle_{\delta q}$ vs. 
the pion mass squared. {\bf[left]} The connected up (octagons) 
and down (diamonds) quark contributions. {\bf[right]} The flavor non-singlet 
$\langle 1 \rangle_{\delta u- \delta d}$.}
\label{fig:1dq_vs_mpi2}
\vspace{-2cm}
\end{figure}
\begin{figure}[t]
\includegraphics[width=7cm]{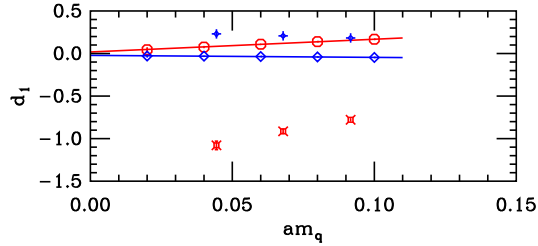}
\vspace{-.5cm}
\caption{The connected $d_1$ matrix element vs. quark mass for the  up (octagons) and down (diamonds) quarks. The up (fancy squares) and down (fancy diamonds) quark for Wilson fermions~\cite{Dolgov:2002zm}.}
\label{fig:d1}
\vspace{-2cm}
\end{figure}
The lowest moment of the transversity $\langle 1 \rangle_{\delta q}$
is also measured. In Fig.~\ref{fig:1dq_vs_mpi2} we plot the
unrenormalized contributions for both the up and down quark, and the
flavor non-singlet combination 
$\langle 1 \rangle_{\delta u - \delta d}$. 
Again the quark mass dependence is very mild and there is no sign of
a chiral log behavior. The ratio 
${\langle 1 \rangle_{\delta u}}/{\langle 1 \rangle_{\delta d}}$ 
linearly extrapolated to the chiral limit 
is also roughly $-4$.

For computing moments of $g_2$ we need to calculate the twist 3 matrix
elements $d_n$. We computed the $d_1$ matrix element which contributes
to the first moment of $g_2$. If chiral symmetry is broken the
operator 
$
 {\cal O}^{[5]q}_{34} = \frac{1}{4}\bar{q}\gamma_5\left[\gamma_3\Dcc_4-\gamma_4\Dcc_3\right]q
$
which is used to compute $d_1$ mixes with the lower dimensional operator 
$
 {\cal O}^{\sigma q}_{34} = \bar{q}\gamma_5\sigma_{34}q.
$ 
Hence in  Wilson fermion calculations a non perturbative
subtraction has to be performed. This has been done for
$d_2$ by QCDSF~\cite{Gockeler:2000ja,Gockeler:2002ek}.  With domain wall
fermions this kind of mixing is proportional to the residual mass 
($\sim\mres/a$),
which in our case is negligible. Thus we expect that a
straightforward computation of $d_1$ with domain wall fermions
provides directly the physically interesting result. In Fig.~\ref{fig:d1}
we present our unrenormalized results for $d_1$ 
as a function of the quark mass.  For
comparison we also plot the unsubtracted quenched Wilson results for
$\beta=6.0$ from~\cite{Dolgov:2002zm}. The fact that our result almost
vanishes at the chiral limit is an indication that the power
divergent mixing is absent for domain wall fermions.
The behavior we find for the $d_1$ matrix element is consistent
with that of the subtracted $d_2$ computed 
by QCDSF~\cite{Gockeler:2000ja,Gockeler:2002ek} with Wilson fermions.
\vspace{-.5cm}

\section{Conclusions}
In conclusion, lattice computations can play an important role in understanding
the hadronic structure and the fundamental properties of QCD. 
Although some difficulties still exist, several significant steps have
been made. Advances in computer technology are expected to play
a significant role in pushing these computations closer to the
chiral limit and in including dynamical fermions. RBC has
already begun preliminary dynamical domain wall fermion 
computations~\cite{Izubuchi:2002pt} which we expect to be pushed forward
with the arrival of QCDOC~\cite{Boyle:2002wg}. 
In the near future, we also expect to complete the
non-perturbative renormalization of the relevant derivative
operators in quenched QCD.
\vspace{-.5cm}


\begin{theacknowledgments} 
I wish to thank Tom Blum, Chulwoo Jung, Shigemi Ohta, and Shoichi
Sasaki for helpful discussions. I also wish to thank the RIKEN BNL
research center, BNL, and the U.S. DOE for providing the facilities
essential for the completion of this work.
\end{theacknowledgments}



\IfFileExists{\jobname.bbl}{}
 {\typeout{}
  \typeout{******************************************}
  \typeout{** Please run "bibtex \jobname" to obtain}
  \typeout{** the bibliography and then re-run LaTeX}
  \typeout{** twice to fix the references!}
  \typeout{******************************************}
  \typeout{}
 }

\end{document}